\documentclass[12pt]{article}
\usepackage{epsfig}
\usepackage{amssymb}
\usepackage{times}
\usepackage{subfigure,wrapfig}
\usepackage{aabib99}
\newcommand{\abb}[1]{Fig.\,\ref{#1}}

\newcommand{\cdr}{\ensuremath{^{13}\textrm{C}}}
\newcommand{\czw}{\ensuremath{^{12}\textrm{C}}}

\newcommand{\nvi}{\ensuremath{^{14}\textrm{N}}}

\newcommand{\ose}{\ensuremath{^{16}\textrm{O}}}

\newcommand{\msun}{\ensuremath{\, {\rm M}_\odot}}

\newcommand{\kelv}{\ensuremath{\,\rm K}}
\newcommand{\jahre}{\ensuremath{\, \mathrm{yr}}}
\newcommand{\mh}{\ensuremath{M_{\rm H}}}        

\newcommand{\mzams}{\ensuremath{M_{\rm ZAMS}}}
\newcommand{\etal}{et al.\,}
\newcommand{\beq}{\begin{equation}}
\newcommand{\beqa}{\begin{eqnarray}}
\newcommand{\eeq}{\end{equation}}
\newcommand{\eeqa}{\end{eqnarray}}
\newcommand{\bedis}{\begin{displaymath}}
\newcommand{\edis}{\end{displaymath}}

\newcommand{\nat}[2]{\ensuremath{#1 \cdot 10^{#2}}}

\newcommand{\apleq}{\ensuremath{\stackrel{<}{_\sim}}}
\newcommand{\apgeq}{\ensuremath{\stackrel{>}{_\sim}}}

\title{\bf Overshoot in Giant Stars
}
\author{F.~Herwig $^1$
and T. Bl\"ocker$^2$
\vspace{1cm}\\
\normalsize $^1$Institut f\"ur Physik, Astrophysik, Universit\"at
Potsdam, Germany\\
\normalsize fherwig@astro.physik.uni-potsdam.de \\
\normalsize $^2$Max-Planck-Institut f\"ur Radioastronomie, Bonn,
Germany\\
\normalsize bloecker@speckle.mpifr-bonn.mpg.de}
\date{\mbox{}}
\begin{document}
\maketitle \pagestyle{empty}
%
%
\def\bull{\vrule height .9ex width .8ex depth -.1ex} \makeatletter
\def\ps@plain{\let\@mkboth\gobbletwo
  \def\@oddhead{}\def\@oddfoot{\hfil\tiny\bull\quad ``The Galactic
    Halo: From Globular Clusters to Field Stars'';
35$^{\mbox{\rm th}}$ Li\`ege\ Int.\ Astroph.\ Coll., 1999\quad\bull}%
\def\@evenhead{}\let\@evenfoot\@oddfoot} \makeatother
%
%
\def\beginrefer{\section*{References}%
\begin{quotation}\mbox{}\par}
\def\refer#1\par{{\setlength{\parindent}{-\leftmargin}\indent#1\par}}
\def\endrefer{\end{quotation}}
%
%
{\noindent\small{\bf Abstract:} The concept of overshoot has already
  been considered for numerous cases in stellar evolution
  calculations. We explore the consequences of overshoot at the
  convection zone which forms during the He-flash (thermal pulse) in
  AGB stars. We find dramatic changes for the abundances within the
  intershell region as well as for the mechanism of the
  3$^\mathrm{rd}$ dredge-up. 
  That means that both the predicted evolution of structural
  quantities as well as the chemical evolution at the surface will be
  different, if overshoot is considered. We also present evidence for
  the presence of overshoot during the He-flash from detailed model
  calculations of the post-AGB phase and the comparison with
  observations. The new evolutionary models
  show that a very late thermal pulse during the post-AGB evolution
  can bring the intershell material up to the surface with minimal
  modification due to convective H-burning. The good agreement of the
  surface abundances of these models with observed surface abundances
  of H-deficient post-AGB stars is interpreted as a strong support for
  the presence of overshoot during the thermal pulses of AGB stars.
       
  }
%
%
\section{Introduction}
The observation of metal-poor stars in globular clusters and field
stars has revealed many well-defined correlations of abundance ratios.
If these results are combined with theoretical
predictions for nucleosynthetic production site the chemical evolution
of stellar populations can be reconstructed. Among the nucleosynthetic
production sites to be 
considered are the asymptotic giant branch (AGB) stars  which are
potentially efficient producers of carbon, nitrogen and $s$-process
elements \cite{jehin:99}. In contrast to the contribution from massive
stars and supernovae, which can be identified down to the lowest
[Fe/H] ratios, AGB stars contribute with some delay to the enrichment
of the ISM.

AGB stars are characterized by an increasingly degenerate C/O core and
two shells of nucleosynthetic processing: the helium burning shell and
the hydrogen 
burning shell \cite{iben:83b}. The He-shell can become thermally
unstable which leads to the recurrent thermonuclear run-away of
He-burning in the shell, the thermal pulses (TP). During the TP the
region between the hydrogen and the helium shell, the intershell
region, becomes convectively unstable due to the huge nuclear energy
production of the He-shell (\abb{KDMOD}) \cite{bloecker:99b}.

The predictions of yields, chemical enrichment and surface abundances
as well as stellar parameters, like luminosity, are sensitively
dependent on the overshoot phenomenon. In the next section we will
review the most important changes introduced to the AGB models if
overshoot is considered. Section \ref{sec:post-AGB} will discuss new
stellar evolution models of hydrogen-deficient post-AGB stars and
their observational counterparts. We argue that this class of stars
can only be understood if overshoot in AGB stars is indeed operating.
\begin{figure}
\centering
  \mbox{\epsfig{file=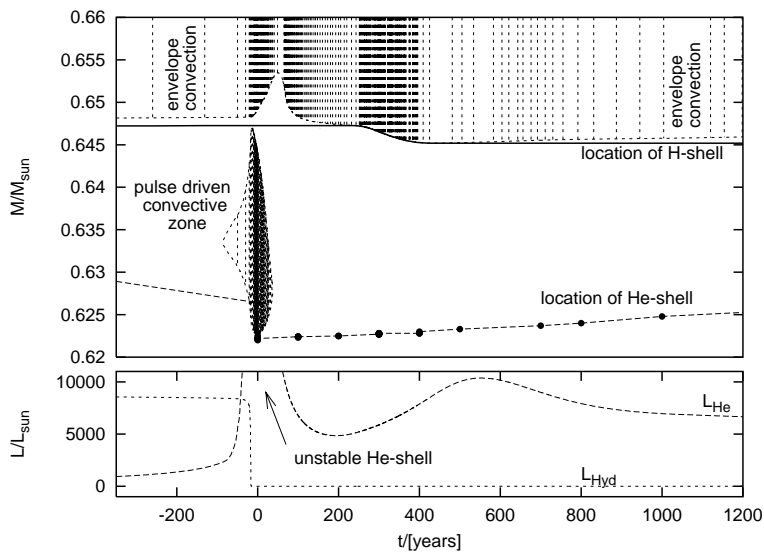,width=10cm}}
\caption{\label{KDMOD}During a thermal pulse.  \textbf{Top panel:}
  Convective regions  
  are indicated by vertical lines, which are spaced according to the
  model density. Around $t=0\jahre$ the He-flash convection zone
  homogenizes the intershell region from the top of the C/O core
  almost up to the mass coordinate of the H-free core. In the
  intershell region the ashes of the H-shell burning accumulate
  during the previous quiescent interpulse phase. This material
  consists mainly of helium ($\simeq 98\%$ by mass) and
  all CNO material has been transformed into \nvi. About $350\jahre$
  after the TP the envelope convection reaches into the intershell
  layer and material previously partially processed by helium burning is
  dredged-up to the surface. \textbf{Lower panel:} Luminosity
  contributions of helium and hydrogen burning. At $t=0\jahre$ the
  peak He-luminosity has been reached. Hydrogen burning has ceased
  because the layers at which hydrogen is still present are
  geometrically lifted due to the expansion work from the
  underlying unstable helium shell. The TP itself has a duration of
  the order of $100 \jahre$.
  }
\end{figure}
\begin{figure}
\centering
  \epsfxsize=0.8\textwidth \epsfbox{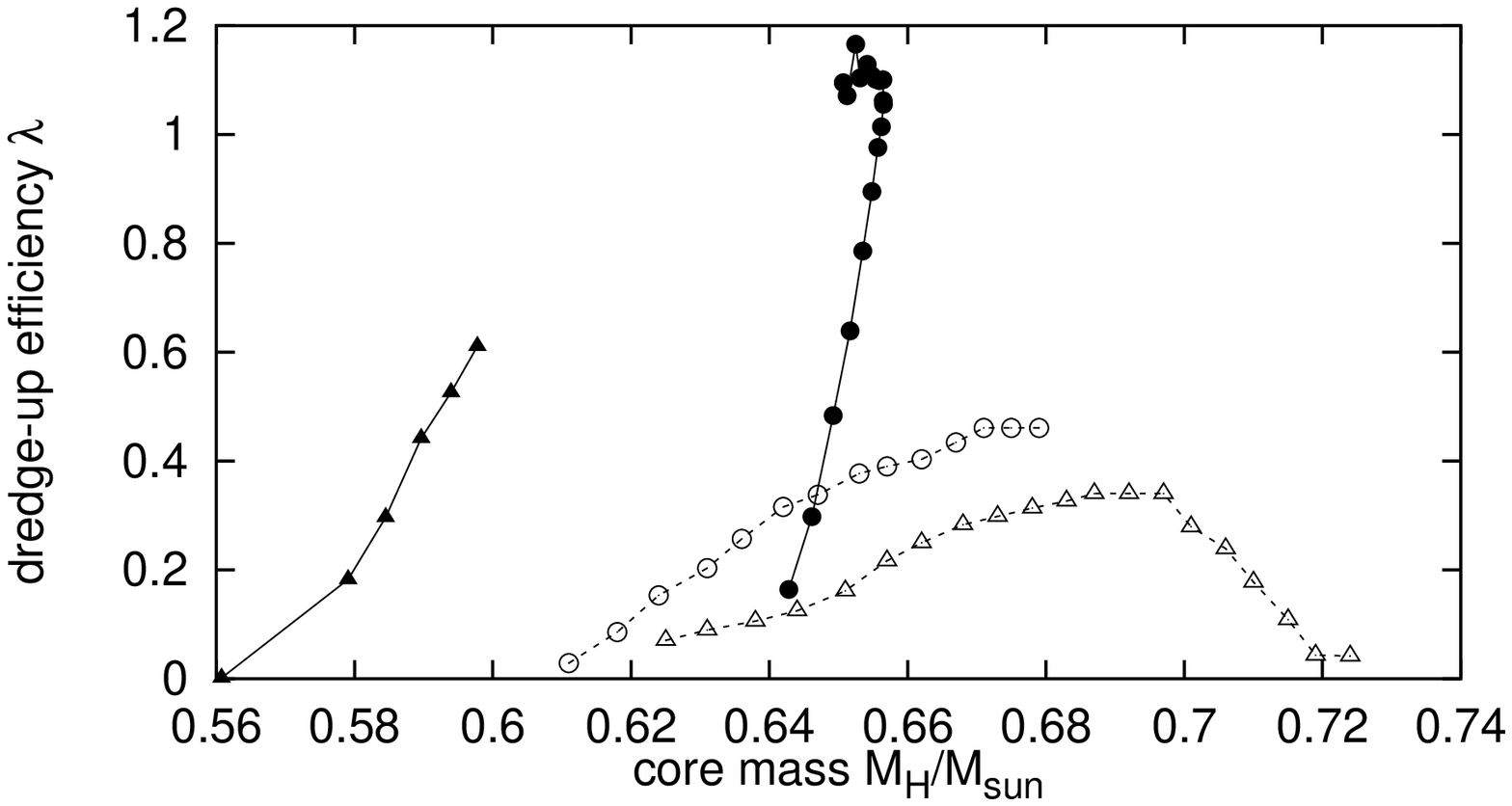}
\caption{\label{dup}
  The dredge-up efficiency $\lambda = \Delta M_\mathrm{DUP}/\Delta
  M\mathrm{H}$ as a function of the core mass for models with
  overshoot (full symbols, this work) and without overshoot (open symbols) by
  Straniero \etal\,(1997). Circles: \mzams=3\msun, full triangles:
  2\msun, open triangles: 1.5\msun. For the 3\msun\ case with
  overshoot the values for 20 TP are shown, of which about 10 are
  clustering at $\lambda \apgeq 1$ because then, by definition, the
  core mass does not grow anymore. All models have solar metallicity  }
\end{figure}

\section{Nucleosynthesis and mixing in AGB stars with overshoot}
Stellar AGB models with overshoot are different from models without
overshoot with respect to
\begin{itemize}
\item the formation of a \cdr\ pocket which is a crucial ingredient of
  the concept of \cdr\ being the dominant neutron source of the main
  solar $s$-process component produced in low-mass AGB stars,
\item the occurrence and efficiency of the third dredge-up
  \cite{herwig:97,herwig:98c},
\item the composition in the intershell and thereby also of the
  ejecta and
\item the structural evolution \cite{herwig:98b}.
\end{itemize}

\subsection{The neutron source for the $s$-process}
At the end of the third dredge-up phase the hydrogen-rich convective
envelope and the carbon-rich intershell region have direct contact.  A
tiny region can form were protons and \czw\ coexist. It has been shown
by Herwig \etal\cite*{herwig:97} that such a region follows naturally
if depth dependent overshoot is considered which models an exponentially
declining turbulent velocity field. During the following interpulse
evolution \cdr\ forms. The \cdr\ burns under radiative conditions
\cite{straniero:95} and releases neutrons via the reaction
$\cdr(\alpha,n)\ose$. Current models of the $s$-process
nucleosynthesis show that this is the dominant mechanism of
heavy-element synthesis in low-mass AGB stars \cite{gallino:97b}.
Together with the H-burning ashes the heavy elements formed by this
mechanism are then  engulfed by the He-flash convection zone and can
reach the surface by the next dredge-up event.

Also note, that dredge-up below the He-flash convection leads to a
 phase where the temperature at the bottom of the
convective boundary exceeds $T \simeq \nat{2.7}{8}\kelv$, which lasts
for about $30 \jahre$ while models without overshoot have a much
shorter high-T phase of only a few years.

\subsection{The third dredge-up}
\label{sec:dup}

\nocite{straniero:97} Maybe one of the most compelling aspects of
models with overshoot is the ease with which they produce low-mass
carbon stars due to efficient dredge-up. For example, our \mzams=1.7\msun,
Z=0.02 model sequence becomes a carbon star after the 10$^\mathrm{th}$
TP when the hydrogen-free core has a mass of $\mh \simeq 0.6\msun$.
The luminosities are $\log L \apleq 4.1$, $\simeq 3.6$ and $\apgeq 3.9$ at
the TP luminosity maximum, minimum and during the quiescent interpulse
phase. Similar results have also been found for models of lower
metallicity and it seems that the problem of the theoretically missing
low-luminosity carbon stars can be solved by some overshoot below the
He-flash convection zone.  The enhanced dredge-up efficiency is caused
by the modification of the intershell abundance.  The abundances found
with overshoot show qualitatively the same dependence on the pulse
number like models without overshoot
\cite{schoenberner:79,boothroyd:88}. However, with overshoot the
intershell contains much more carbon and oxygen at the expense of
helium. For example, for a 3\msun\ stellar model
after the 13$^\mathrm{th}$ TP we find
[He,C,O]=[0.39,0.41,0.16]\footnote{All abundances are given in mass
  fractions.} whereas the values from calculations without overshoot
are [He,C,O]=[0.70,0.26,0.01]. It has actually been found that the
amount of overshoot applied to the bottom of the He-flash convection
zone, the resulting helium abundance and the dredge-up after the TP
are proportional \cite{herwig:98c}.

The combined effect of efficient dredge-up  and enhanced
oxygen and carbon content in the intershell region inverts the common
belief that AGB stars do not contribute to the Galactic oxygen
production. Models with overshoot predict that a 3\msun\ star of solar
metallicity will eject about 
0.03\msun\ of primary \ose\ which was formed by He-shell
burning. Also, the surface abundance of oxygen (and thus the O/H
ratio)  is larger by about a factor
of two to three due to overhoot. Unfortunately, the 
determination of the oxygen abundance in giant stars is difficult and
yields sometimes ambiguous results \cite{fulbright:99}.

\section{Evidence for overshoot in AGB stars from models and
  observations of post-AGB stars}
\label{sec:post-AGB}

  \begin{figure}
    \centering
    \subfigure[Without overshoot]{\epsfig{figure=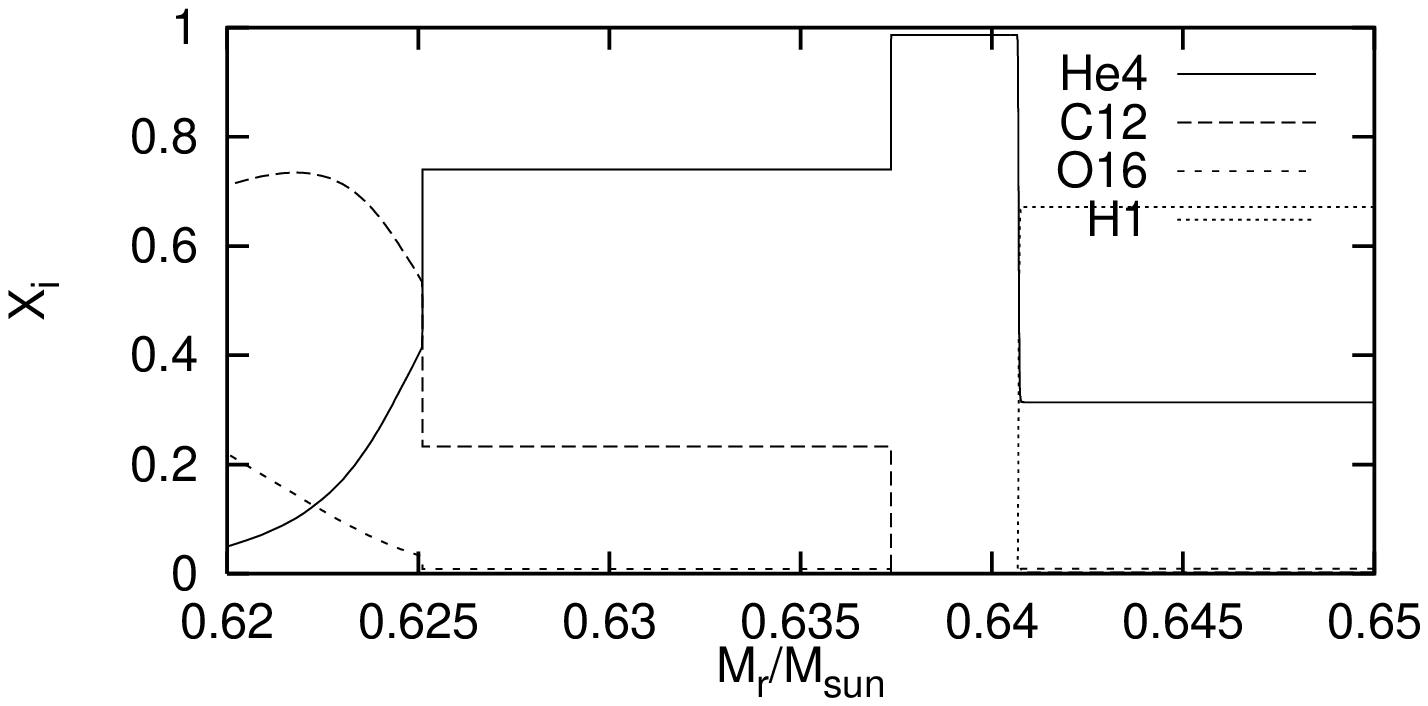,width=0.47\textwidth}}
    \subfigure[With overshoot]{\epsfig{figure=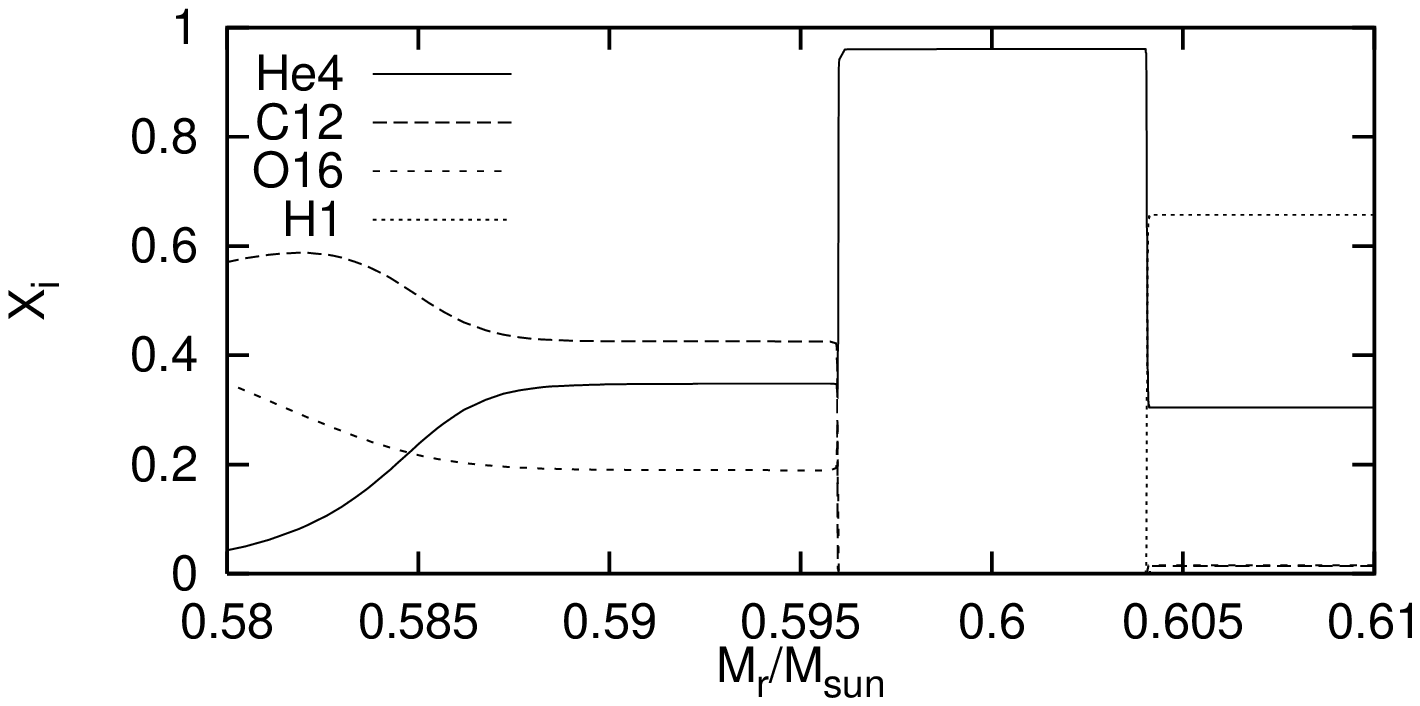,width=0.47\textwidth}}
    \caption{Abundance profile of AGB models in the region of the He-
    and H-burning shell.}
    \label{profiles}
  \end{figure}
\begin{figure}
\centering
  \epsfxsize=0.8\textwidth
  \epsfbox{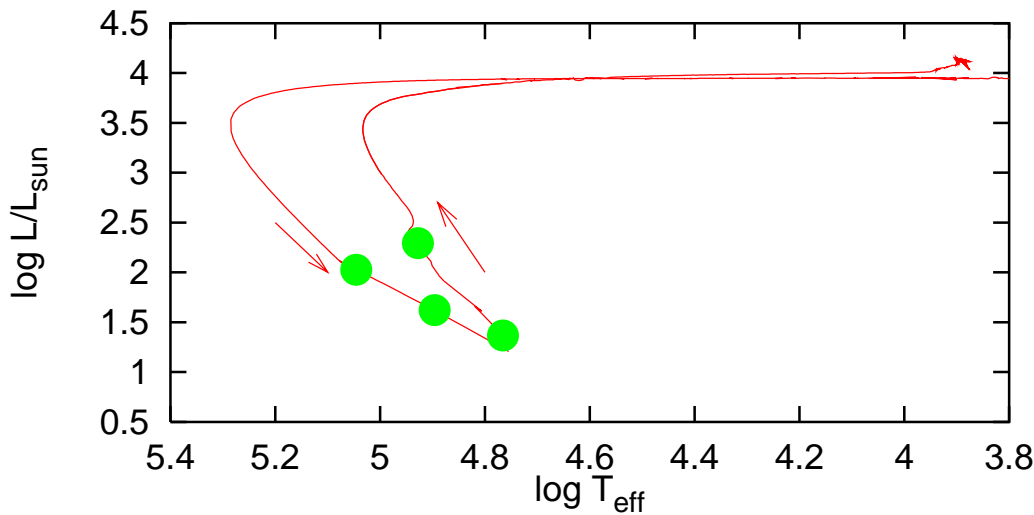} 
\caption{\label{HRD}
  Post-AGB evolution with a very late thermal pulse and consecutive
  return into the AGB domain of the HRD.
  At the first mark
  along the track, the He-flash convection zone is already well
  established.
  At the second mark, the outwards
  growing convection zone has reached the envelope and protons start
  to enter into the convective zone were they are captured by \czw\
  \emph{on the way}.
  At the third mark the hydrogen
  luminosity has reached its peak. The surface composition is hydrogen-free
  beyond the last mark. }
\end{figure}
  In the previous section we have shown that overshoot from below the
He-flash convection zone leads to important modifications of structure 
and chemical evolution of AGB stars. It is therefore important to
collect evidence for the existence and efficiency of overshooting at this 
convective boundary. Such evidence comes from a class of H-deficient post-AGB
stars with unusual surface abundances: the [WC]-type central stars of 
planetary nebulae \cite{koesterke:97b,hamann:97,demarco:97} and the
PG\,1159 stars \cite{dreizler:98,werner:98}. 
The spectroscopic abundance analysis revealed that these
stars are very carbon rich and also \emph{oxygen rich!}. Typically,
one finds [He,C,O]=[0.50,0.33,0.17] for PG\,1159 stars. In particular
the large mass 
fraction of oxygen has been puzzling because AGB stellar models have
shown such a high oxygen abundance only \emph{below} the intershell and
helium-burning region where the helium abundance is already much lower 
then the observed $50\%$. Thus, AGB models without overshoot do not
show the observed abundance pattern of H-deficient post-AGB stars at
any depth  and, moreover, no scenario of combined mixing and mass
loss could be identified to solve the problem. Almost any post-AGB
evolutionary model sequence is H-normal \cite{bloecker:95b}.
The only advance has
been the discovery that a thermal pulse on the post-AGB, shortly before 
the star becomes a white dwarf, may lead to convective H-burning in
the intershell during the He-flash (see Iben \& McDonald
\cite*{iben:95b} and references there). This \emph{very late}
TP brings up the intershell abundance and thereby leads to a
significant abundance change at the surface. However, the models
which where available so far fail to predict the observed high
abundance of oxygen.  

With  overshoot the abundances in the
intershell coincide naturally with the abundance pattern found in the 
H-deficient post-AGB stars (\abb{profiles}). Also, a certain range of observed
abundances may reflect the time  and
metallicity dependence of the intershell abundance. We have performed
new stellar evolution calculations of the post-AGB phase
which take the modifications of the 
intershell abundance of AGB stars due to overshoot into account. Also, 
we employ  a time-dependent and coupled treatment of 
nuclear burning and convective mixing to correctly describe the
nuclear processing of protons under the conditions encountered in the
convectively unstable He-shell (for details see Herwig
\cite*{herwig:00a}).
We found that, within the very late
thermal pulse scenario, H-deficient Born-again post-AGB stars (see \abb{HRD}) exhibit
surface abundances (also for oxygen) and parameters in the range
observed \cite{herwig:99c}.

\section{Conclusions}
In this paper we have reported of interdependent
advances of recent stellar evolution modeling of AGB and
post-AGB stars which mutually add to their significance. Currently the
modification of the intershell abundances following the overshoot
concept appears to be the \emph{only} possible theoretical
explanation for the large oxygen abundances observed in H-deficient
post-AGB stars. Thus, the application of some overshoot from below the 
He-flash convection zone of AGB stars solves a long standing
observational puzzle.

In turn, this importance of AGB-overshoot for the H-deficient
post-AGB stars presently is one of the strongest evidence that such
overshoot is indeed operating. Also other problems can  be
resolved, like the modeling of carbon stars with low luminosities as
observed in the Magellanic Clouds. More support for the overshoot
concept comes also from hydrodynamic modeling (see contribution by
D.\,Arnett, this volume). On the other hand, we have so far not found 
any evidence against the operation of overshoot in AGB stars (though
we might not have searched for it hard enough yet).

At the present stage we can conclude that the operation of some
overshoot in AGB stars appears to be very likely. The consequences for 
the chemical yields and the enrichment are both qualitative and
quantitative. While the larger dredge-up efficiency enhances the mass
fraction of synthesized material in the ejecta in general also the
relative abundance fractions are changed, which is most remarkably the
case for oxygen.

%
%
\section*{Acknowledgements}
This work has been supported by the \emph{Deut\-sche
  For\-schungs\-ge\-mein\-schaft, DFG\/} (La\,587/16).
We would also like to thank Drs.\ W.-R.\, Hamann, L.\, Koesterke and N.\,
Langer for very helpful discussion.     
%
%


\begin{thebibliography}{}

\bibitem[\protect\astroncite{Bl\"ocker}{1995}]{bloecker:95b}
Bl\"ocker, T., 1995,
\newblock {A\&A} {299}, 755

\bibitem[\protect\astroncite{Bl\"ocker}{1999}]{bloecker:99b}
Bl\"ocker, T., 1999,
\newblock in T.~L. Bertre, A. Lebre, and C. Waelkens (eds.), {AGB Stars}, IAU
  Symp.\,191, p.~21, PASP

\bibitem[\protect\astroncite{Boothroyd and Sackmann}{1988}]{boothroyd:88}
Boothroyd, A.~I. and Sackmann, I.-J., 1988,
\newblock {ApJ} {328}, 671

\bibitem[\protect\astroncite{{De Marco} et~al.}{1998}]{demarco:97}
{De Marco}, O., {Storey}, P.~J., and {Barlow}, M.~J., 1998,
\newblock {MNRAS} {297}, 999

\bibitem[\protect\astroncite{Dreizler and Heber}{1998}]{dreizler:98}
Dreizler, S. and Heber, U., 1998,
\newblock {A\&A} {334}, 618

\bibitem[\protect\astroncite{{Fulbright} and {Kraft}}{1999}]{fulbright:99}
{Fulbright}, J.~P. and {Kraft}, R.~P., 1999,
\newblock {AJ} {118}, 527

\bibitem[\protect\astroncite{Gallino et~al.}{1998}]{gallino:97b}
Gallino, R., Arlandini, C., Busso, M., Lugaro, M., Travaglio, C., Straniero,
  O., Chieffi, A., and Limongi, M., 1998,
\newblock {ApJ} {497}, 388,
\newblock in press

\bibitem[\protect\astroncite{Hamann}{1997}]{hamann:97}
Hamann, W.-R., 1997,
\newblock in H. Habing and H. Lamers (eds.), {Planetary Nebulae}, Vol. IAU
  Symp.\,180, p.~91, Kluwer

\bibitem[\protect\astroncite{Herwig}{2000}]{herwig:00a}
Herwig, F., 2000,
\newblock in T. Bl\"ocker, R. Waters, and B. Zijlstra (eds.), {Low mass
  Wolf-Rayet Stars: origin and evolution}, Ap\&SS, Kluwer,
\newblock in press

\bibitem[\protect\astroncite{Herwig et~al.}{1999a}]{herwig:99c}
Herwig, F., Bl\"ocker, T., Langer, N., and Driebe, T., 1999a,
\newblock {A\&A} {349}, L5

\bibitem[\protect\astroncite{Herwig et~al.}{1999b}]{herwig:98c}
Herwig, F., Bl\"ocker, T., and Sch\"onberner, D., 1999b,
\newblock in T.~L. Bertre, A. Lebre, and C. Waelkens (eds.), {AGB Stars},
  p.~41, PASP

\bibitem[\protect\astroncite{Herwig et~al.}{1997}]{herwig:97}
Herwig, F., Bl\"ocker, T., Sch\"onberner, D., and {El Eid}, M.~F., 1997,
\newblock {A\&A} {324}, L81

\bibitem[\protect\astroncite{Herwig et~al.}{1998}]{herwig:98b}
Herwig, F., Sch\"onberner, D., and Bl\"ocker, T., 1998,
\newblock {A\&A} {340}, L43

\bibitem[\protect\astroncite{Iben and McDonald}{1995}]{iben:95b}
Iben, Jr., I. and McDonald, J., 1995,
\newblock in D. Koester and K. Werner (eds.), {White Dwarfs}, No. 443 in LNP,
  p.~48, Springer, Heidelberg

\bibitem[\protect\astroncite{Iben and Renzini}{1983}]{iben:83b}
Iben, Jr., I. and Renzini, A., 1983,
\newblock {ARA\&A} {21}, 271

\bibitem[\protect\astroncite{Jehin et~al.}{1999}]{jehin:99}
Jehin, E., Magain, P., Neuforge, C., Noels, A., Parmentier, G., and Thoul,
  A.~A., 1999,
\newblock {A\&A} {341}, 241

\bibitem[\protect\astroncite{{Koesterke} and {Hamann}}{1997}]{koesterke:97b}
{Koesterke}, L. and {Hamann}, W.~R., 1997,
\newblock {A\&A} {320}, 91

\bibitem[\protect\astroncite{Sch\"onberner}{1979}]{schoenberner:79}
Sch\"onberner, D., 1979,
\newblock {A\&A} {79}, 108

\bibitem[\protect\astroncite{Straniero et~al.}{1997}]{straniero:97}
Straniero, O., Chieffi, A., Limongi, M., Busso, M., Gallino, R., and Arlandini,
  C., 1997,
\newblock {ApJ} {478}, 332

\bibitem[\protect\astroncite{Straniero et~al.}{1995}]{straniero:95}
Straniero, O., Gallino, R., Busso, M., Chieffi, A., Raiteri, C.~M., Salaris,
  M., and Limongi, M., 1995,
\newblock {ApJ} {440}, L85

\bibitem[\protect\astroncite{Werner et~al.}{1999}]{werner:98}
Werner, K., Dreizler, S., Rauch, T., Koesterke, L., and Heber, U., 1999,
\newblock in T.~L. Bertre, A. Lebre, and C. Waelkens (eds.), {AGB Stars}, IAU
  Symp.\,191, p. 493, PASP

\end{thebibliography}
\end{document}